\documentclass[pdftex,a4paper,11pt]{article}

\usepackage[pdftex]{graphicx}
\usepackage{amssymb}
\usepackage[thickspace,thickqspace,amssymb]{SIunits}
\usepackage[margin=2.5cm]{geometry}
\usepackage{cmbright}
\usepackage[small,bf]{caption}
\usepackage{authblk}

\usepackage[T1]{fontenc}

\begin{document}

\title{The Cosmic Ray p+He energy spectrum in the 3-3000 TeV energy range measured by ARGO-YBJ}

\author[1]{S. M. Mari}
\author[2]{P. Montini}
\affil[ ]{on behalf of the ARGO--YBJ Collaboration}
\affil[1]{Dipartimento di Matematica e Fisica, Universit\`a degli Studi Roma Tre}
\affil[2]{Istituto Nazionale di Fisica Nucleare (INFN) - Sezione di Roma Tre}

\maketitle

\abstract{The ARGO--YBJ  experiment is a full coverage air shower detector operated at the Yangbajing International Cosmic Ray Observatory. The detector has been in stable data taking in its full configuration since November 2007 to February 2013. The high altitude and the high segmentation and spacetime resolution offer the possibility to explore the cosmic ray energy spectrum in a very wide range, from a few TeV up to the PeV region. The high segmentation allows a detailed measurement of the lateral distribution, which can be used in order to discriminate showers produced by light and heavy elements. In this work we present the measurement of the cosmic ray light component spectrum in the energy range 3--3000 TeV. The analysis has been carried out by using a two-dimensional unfolding method based on the Bayes' theorem.}

\section{Introduction}
The Earth is permanently exposed to a flux of cosmic rays coming from outside the solar system. Their energy spectrum spans a huge energy range (from MeV up to $\sim 10^{20}$ eV), corresponding to a variation of the intensity of several orders of magnitude. The all-particle energy spectrum of cosmic rays can be roughly described by a power--law, showing a  ``knee'' at energies around 4 PeV. Despite great experimental efforts in the last decades, the origin of the knee is still an unresolved question. In the standard picture the knee in the all particle spectrum is due to the steepening of the proton and helium spectra. Due to the rapid decrease of the intensity as the energy increases, all information about cosmic rays above 100 TeV comes from ground--based air shower experiments and the determination of the composition is essentially limited to large mass groups. A precise measurement of the proton and helium spectra in the energy region from few TeV to 10 PeV is of fundamental importance in the understanding of the origin of the knee. The ARGO--YBJ experiment is a full--coverage extensive air shower array which was in stable data taking from November 2007 to February 2013. The detector is equipped with two independent readout systems designed in order to detect showers in a very wide range of particle density. The full--coverage technique, combined with the high segmentation and spacetime resolution, allows a detailed measurement of the lateral distribution of the shower that can be used in order to discriminate events produced by light primaries.

\section{The ARGO--YBJ experiment}
\label{argo}
The ARGO--YBJ experiment \cite{argo1,argo2} is a full coverage air shower detector operated in the Yangbajing International Cosmic Ray Observatory (Tibet, P.R. China, 4300 m a.s.l.). The detector has been designed and developed in order to explore the cosmic ray energy spectrum and composition in a wide energy range from few TeV up to the PeV region. The detector is made of a single layer of 1836 resistive plate chambers (RPCs)  with $\sim 93 \%$ active area and equipped with two independent readout systems. The carpet is made of 1836 RPCs, arranged in 153 clusters each made of 12 chambers.  The digital readout consists of 146880 copper strips ($6.65 \times 61.80 \, \mathrm{cm^2}$) logically arranged in 18360 pads. The high segmentation and spacetime resolution of the digital readout allow the detection of showers produced by low energy primaries down down to few TeV and provide a detailed reconstruction of showers up to $\sim 23\, \mathrm{particles/m^2}$ corresponding to a primary energy up to a few hundreds of TeV. In order to extend the operating range of the detector and fully investigate the PeV energy region, each RPC has been equipped with two large electrodes called Big Pads ($139 \times 123 \, \mathrm{cm^2}$), each providing a signal that is proportional to the number of charged particles impinging the detector surface.  The analog readout system uses eight different gain scales (G0, $\ldots$, G7) giving an overlap between the analog and digital readout linearity range. Data from the highest gain scale (G7) have been used for calibration purposes. The intermediate gain scale (G4) overlaps with the digital readout data in a wide energy range between 10 and 100 TeV, providing a cross--calibration of the two techniques. Data from the lowest gain scales (G1 and G0) allow the detection of showers with about $10^4$ particles/$\mathrm{m^2}$ in the core region. A dedicated calibration procedure has been implemented for each gain scale \cite{michele-calib}. 

\section{Data analysis}

As it is well known, since the development of a shower presents intrinsic fluctuations, the primary energy cannot be determined on an event--by--event basis. As widely described in \cite{bayes,light12}, the determination of the cosmic ray energy spectrum from the spacetime distribution of charged particles at ground level is a classical unfolding problem that can be dealt with the bayesian technique \cite{dagos}. In order to evaluate the conditioned probabilities needed in the unfolding procedure a detailed simulation of the development of the shower in the Earth's atmosphere and of the detector response has been performed. A sample of simulated events was generated by using the CORSIKA code \cite{corsika}, including FLUKA and QGSJETII.03 hadronic interaction models. Showers produced by protons, helium, CNO and iron nuclei have been generated in the energy range $(0.316 - 31600)\, \mathrm{TeV}$ and in the zenith angle range $0-45$ degrees according to the energy spectra described in \cite{horandel}. Both the digital and analog output of the detector have been simulated by using a GEANT3 based code, including effects of time resolution, RPC efficiency, noise, etc. 

\section{The cosmic ray p+He spectrum}

\subsection{3--300 TeV energy range}
\label{digital}

\begin{figure}
\begin{center}
\includegraphics[width = 0.9\textwidth]{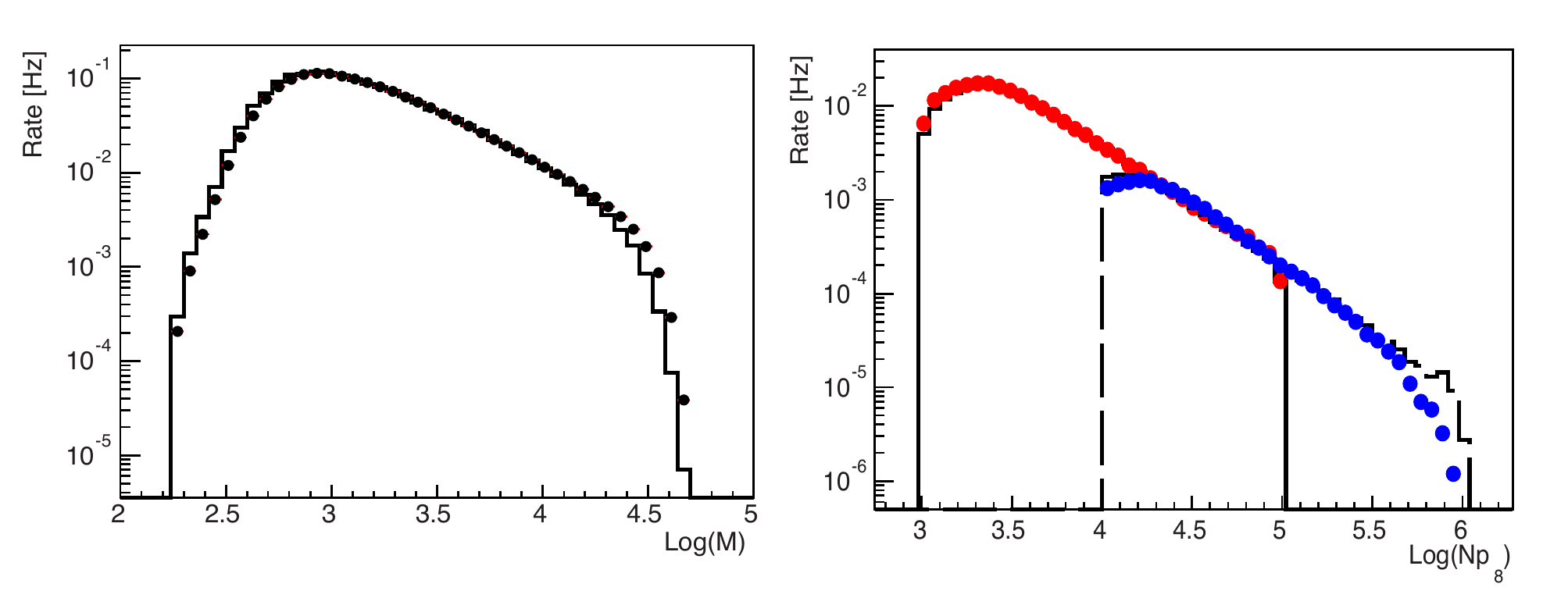}
\caption{Left: multiplicity distribution of events selected according to the criteria described in section \ref{digital}. Values for Monte Carlo (line) and experimental data (dots) are reported. Right: distribution of $Np_{8}$ of events selected according to the criteria described in section \ref{sect:analog}. Values for G4 Monte Carlo (solid line), G4 data (red dots), G1 Monte Carlo (dashed line) and G1 data (blue dots) are reported. }
\label{mult}
\end{center}
\end{figure}

The digital readout of the ARGO--YBJ experiment has been operated in its full configuration for more than five years and more than $5\times 10^{11}$ events have been recorded and reconstructed. A first selection has been based on the working condition of the detector and on the quality of the reconstruction procedure in order to obtain a sample of high--quality runs. About $3\times 10^{11}$ events have been selected. In order to obtain a better estimation of the bayesian probabilities needed in the unfolding procedure the following selection criteria have been applied to both Monte Carlo and experimental data samples.
\begin{itemize}
\item Events with reconstructed zenith angle in the range $0^\circ \leqslant \vartheta_{R} \leqslant 35^\circ$ have been considered.
\item The measured strip multiplicity $M$ had to be in the range $150 \leqslant M \leqslant 50000$.
\item The position of the cluster with the highest multiplicity must be located inside an area of $40 \times 40 \, \mathrm{m^2}$ measured from the detector center
\item The particle density measured by the innermost 20 clusters $\rho_{in}$ must be higher than the one measured by the outermost 42 ones ($\rho_{out}$). 
\end{itemize}
These selection criteria reduces bias effects in the estimation of the probabilities needed in the unfolding procedure. The particle density cut, moreover, allows the selection of showers manly produced by light primaries. In figure \ref{mult} the strip multiplicity distribution is reported for both data and Monte Carlo samples. The analysis has been performed on the full sample collected during the period 2008--2012 and the resulting spectrum is reported in figure \ref{multitev}.  Special care has been dedicated to the evaluation of the possible systematic effects. The main sources of systematic uncertainties rely on the reliability of the detector simulation, effects related to the values of the cuts used in the event selection procedure and the contamination of heavier elements. The total systematic uncertainty has been estimated as of the order of $\pm5\%$. The ARGO--YBJ measurement spans a wide energy range, from a few TeV up to hundreds of TeV, so far accessible only by balloon--borne or satellite experiments.

\subsection{10--3000 TeV energy range}
\label{sect:analog}

\begin{figure}[h]
\begin{center}
\includegraphics[width = 0.9\textwidth]{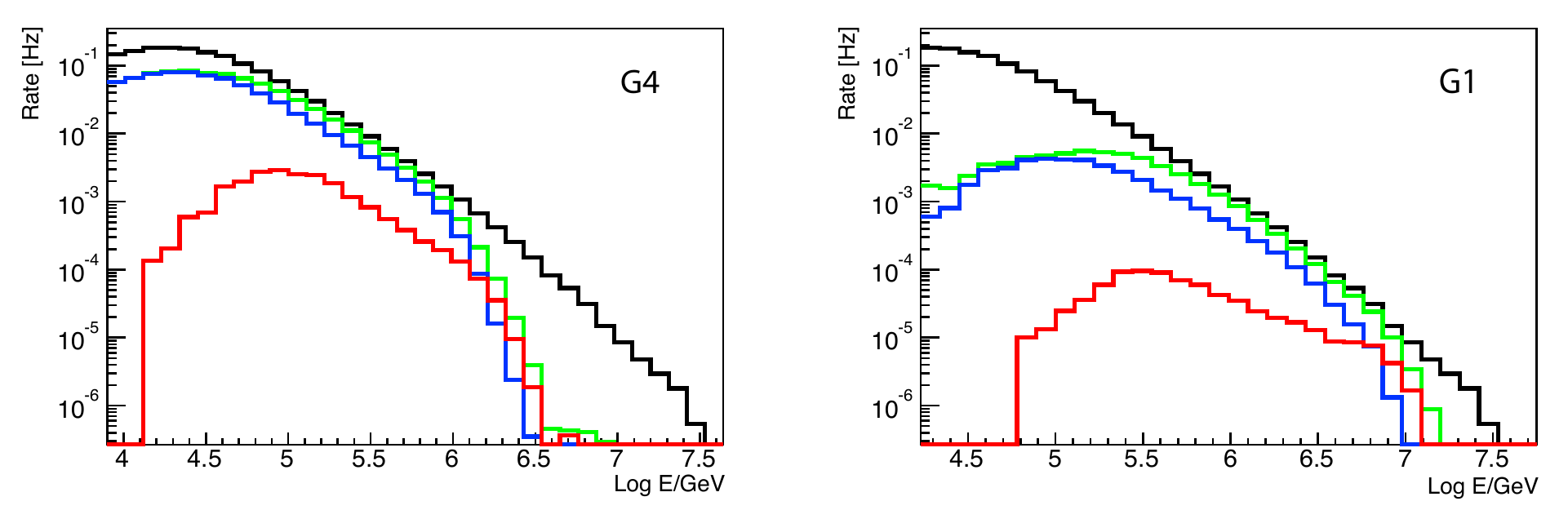}
\caption{Energy distribution of Monte Carlo events surviving the fiducial cuts described in section \ref{sect:analog} (green). Energy spectra of the light (blue) and heavy (red) component after the composition cut are also reported. The simulated spectrum is reported as a reference pattern (black). }
\label{selG1G4}
\end{center}
\end{figure}

The measurement of the spectrum up to 3 PeV has been carried out by exploiting the RPC analog readout system of the ARGO--YBJ experiment. As a first step, several studies have been performed on the simulated data sample in order to identify a suitable energy estimator and a set of parameters that allow the discrimination between showers produced by light and heavy primaries respectively. The number of particles within 8 m from the core position ($Np_{8}$) seems to be well correlated with primary energy and is not affected by bias effects related to the finite detector size. The lateral distribution of charged particles have been studied at several distances from the core position with high precision. Showers produced by light primaries have a well--shaped core and the largest fraction of particles is localized around the core region. Showers generated by heavy nuclei, otherwise, show larger particle density at distances far from the core. The ratio $\beta$ between the particle density at several distances from the core position and the particle density measured in the core region can therefore be exploited in order to discriminate light primaries. In this work we present the analysis of the first data collected during 2010. The analysis has been performed on the data collected by using the G4 and G1 gain scales. The following fiducial cuts have been applied to both Monte Carlo and experimental data samples:
\begin{itemize}
\item showers with reconstructed zenith angle in the range $0^\circ \leqslant \vartheta_{R} \leqslant 35^\circ$ have been considered,
\item reconstructed core position located inside an area of $40 \times 40\, \mathrm{m^2}$ around the detector center,
\item $Np_{8}$ must be in the range $(10^3-10^5)$ for G4 and $(10^4-10^6)$ for G1 data samples.
\end{itemize}

\begin{figure}
\begin{center}
\includegraphics[width = 0.9\textwidth]{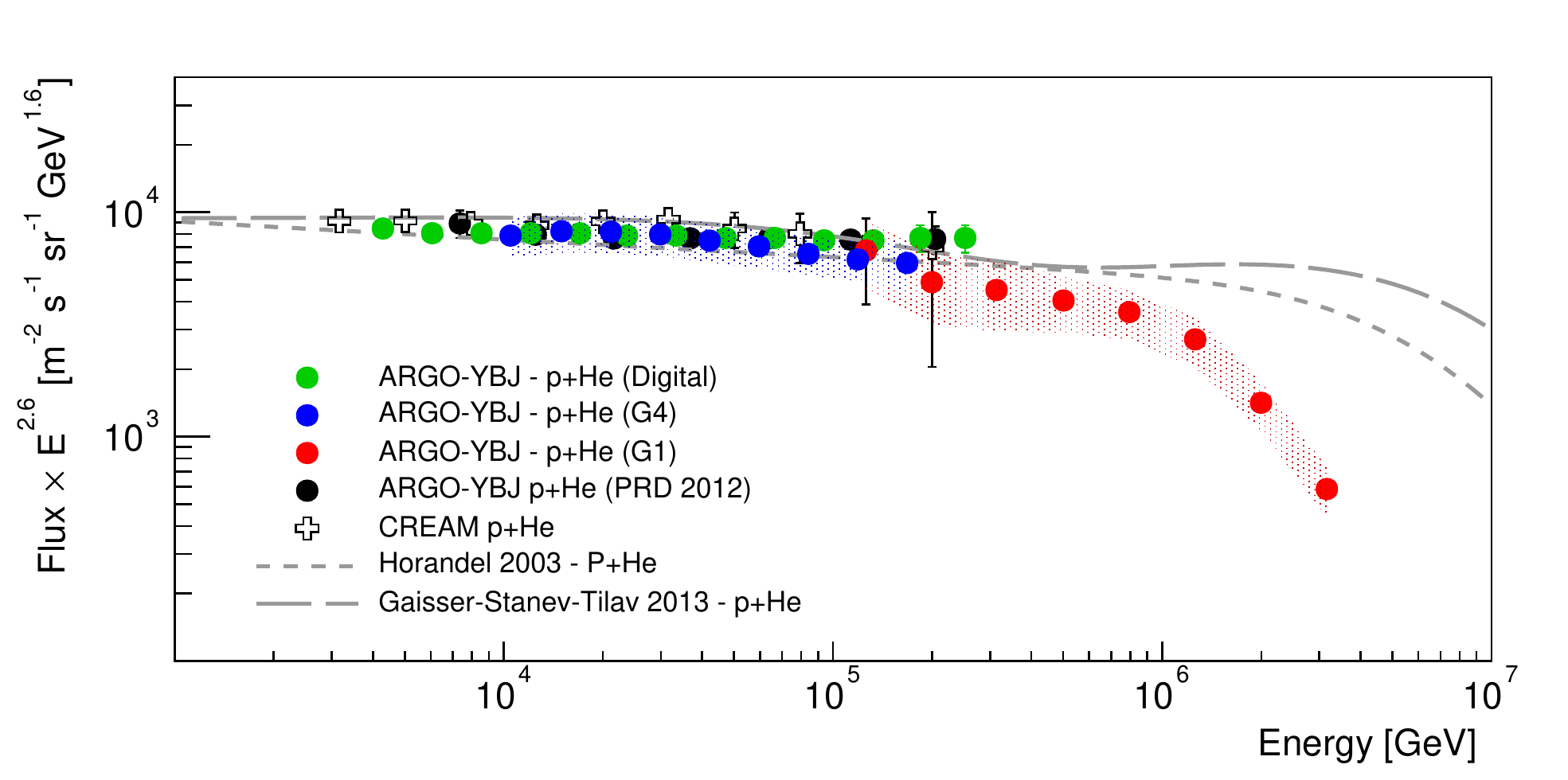}
\caption{The p+He spectrum measured by the ARGO--YBJ experiment by analyzing the full digital data sample (green dots). Preliminary results of G4 (blue dots) and G1 (red dots) analysis are reported. The shaded area around G4 and G1 values represents the systematic uncertainty.  The measurement obtained by analyzing the first data of ARGO--YBJ is also reported \cite{light12}. The sum of the proton and helium spectra measured by CREAM \cite{cream} is shown (crosses). The curves of the p+He spectrum according to the H\"orandel (solid line) \cite{horandel} and GST (dashed line) \cite{GST} models are also shown.}
\label{multitev}
\end{center}
\end{figure}

In figure \ref{mult} the distribution of $Np_8$ is reported for both Monte Carlo and data samples, demonstrating the reliability of the simulation of the analog readout output. An additional selection criterion based on $\beta$ has been applied to the events surviving the fiducial cuts. In figure \ref{selG1G4} the fraction of events surviving the fiducial cuts and the discrimination cut is presented for both G1 and G4 data samples. The plot shows clearly that the selected samples are essentially made of showers produced by light primaries. The G4 and G1 data sets have been analyzed in order to measure the $N(Np_8)$ distribution, while the Monte Carlo data sample has been used in order to evaluate the probabilities needed in the unfolding procedure. The resulting spectra are shown in figure \ref{multitev}. The energy spectrum obtained by analyzing the G4 data sample spans the energy range between 10 and 150 TeV, overlapping the spectrum measured by using the digital readout data as described in section \ref{digital}. These results are fairly consistent between each other, both concerning the spectral index and flux intensity, demonstrating the reliability of the response of the analog readout system. The analysis of the G1 dataset extend the energy range up to the PeV region, clearly showing a deviation from a single power law at energies of about 700 TeV.   The same sources of systematic effects mentioned in section \ref{digital} have been considered.

\section{Conclusions}
The ARGO--YBJ experiment has been taking data in its full configuration for more than five years. The detector has been equipped with two independent readout systems. The high segmentation and spacetime resolution of the digital readout system allows the detection of showers produced by primaries in a wide energy range, from a few TeV up to a few hundreds of TeV. The analog readout system has been designed and implemented in order to sample showers with about $10^4\, \mathrm{particles/m^2}$ around the core region, corresponding to a primary energy up to the PeV region. Three independent datasets have been analyzed in order to investigate the cosmic ray energy spectrum in the region $(3-3000)\, \mathrm{TeV}$.  The relation between the shower size and the primary energy has been established by using a bayesian unfolding procedure.  The discrimination between showers produced by light and heavy primaries has been obtained by using a selection criterion based on the lateral distribution of charged particles in the shower front. The analysis of the digital data sample collected during the $\sim 5$ years lifespan of the experiment has been performed. The resulting spectrum spans the energy range $(3-300) \, \mathrm{TeV}$, giving a spectral index $\gamma = -2.62 \pm 0.01$, in excellent agreement with the one obtained by analyzing the first data taken with the detector in its full configuration \cite {light12}. The analysis of the analog readout data sample has been performed in order to measure the p+He spectrum in the energy region $(10-3000) \, \mathrm{TeV}$ energy range. The resulting spectrum remarkably agrees with the one obtained by analyzing the digital readout full data sample in a wide energy range from 10 TeV up to 150 TeV. A deviation from a single power law is clearly evident at energies of about 700 TeV, showing a rapid decrease of the light component flux at higher energies. These results demonstrate the possibility of exploring the cosmic ray properties in a wide energy range with a single ground--based experiment and open new scenarios about the evolution of the light component energy spectrum towards the highest energies and about the origin of the knee.

\end{document}